\documentstyle[pra,aps,epsf,twocolumn]{revtex} 

\title{Threshold and linewidth of a mirrorless parametric oscillator}

\author{M.~Fleischhauer$^{1,2}$, M.~D.~Lukin$^{1}$,  A.~B.~Matsko$^{3,4}$, and 
M.~O.~Scully$^{3,4}$}
\address{$^1$ ITAMP, Harvard-Smithsonian
Center for Astrophysics, Cambridge, MA 02138,}
\address{$^2$Sektion Physik,  
Universit\"at M\"unchen, 
D-80333 M\"unchen, Germany}
\address{$^3$Department of Physics, Texas A\&M University,
         College Station, Texas 77843-4242,}
\address{$^4$Max-Plank-Institut f\"ur Quantenoptik, Garching, D-85748, Germany,}

\date{January 31}
  
\begin{document}

\maketitle

\begin{abstract}
We analyze the above-threshold behavior of a
mirrorless parametric oscillator based on resonantly enhanced
four wave mixing in a coherently driven dense atomic vapor.  
It is shown that, in the ideal limit, an 
arbitrary small flux of pump photons is sufficient
to reach the oscillator threshold. We demonstrate that due to the large 
group velocity delays associated with coherent media,
an extremely narrow oscillator linewidth is possible, making a
narrow-band source of non-classical radiation feasible. 
\end{abstract}


\tighten

Stable and low-noise sources of coherent and non-classical radiation are 
of interest in many areas of laser physics and quantum optics. Such 
sources have a  wide range of 
applications such as frequency standards, optical magnetometry, 
gravitational wave detection, and high-precision spectroscopy.  

The present theoretical work is motivated by
recent experiments demonstrating a phase transition to
mirrorless oscillation of counter-propagating Stokes and 
anti-Stokes  fields in resonant,  double-$\Lambda$ Raman media 
\cite{zibrov98}. In contrast to earlier studies involving instabilities
in alkali vapors \cite{old1,old2,old3,old4}, 
this oscillation could be achieved with 
pump fields of $\mu$W power (nano Joule pulse energy) and is 
accompanied by a dramatic  narrowing of the beat signals between 
driving and generated fields. Oscillations of this kind 
are a clear manifestation of atomic coherence and interference effects, which 
have recently lead to many exciting developments in resonant nonlinear 
optics \cite{hemmer95,harris97phys.today,jain96,rev99}. In particular, 
the unusual efficiency of the present processes is expected to lead to
a new regime of quantum nonlinear optics in which interactions at a 
level of few light quanta  are feasible. Furthermore, 
the photon pairs generated can possess nearly ideal quantum 
correlations, resulting in almost  complete squeezing of quantum fluctuations
\cite{lukin99}.

We here study theoretically the quantum dynamics of 
the mirrorless oscillator above threshold.
We show that for an infinitely long lived atomic dark state
an arbitrary small stationary flux of pump photons is sufficient to
maintain the oscillation. 
We furthermore analyze frequency locking and linewidth narrowing 
of the beat note between oscillation-  and pump frequencies. 
In particular, we show that the   
beat-note linewidth is given by
an expression similar to the Schawlow-Townes formula for lasers
where the cavity storage time is replaced by the group delay time
$\tau_{gr}$ in the medium.
Due to the large linear dispersion associated with
electromagnetically induced transparency (EIT) in optically
thick media, the group delay can be
extremely large 
\cite{hau99,Schmidt96,group_delay_preprint} leading 
to a very small beat-note linewidth. This effect is analogous to the 
line-\-narrowing in intracavity  EIT 
\cite{lukin98ol,fleischhauer94}. Since
only very small pump powers are needed to reach threshold, ac-Stark shifts
and the associated systematic effects on the beat-note frequency
can be made very small. 
The combination of line-narrowing and small pump-power
requirements makes the mirrorless 
parametric 
oscillator 
an interesting novel source of stable and narrow-linewidth non-classical 
radiation.
Possible applications include frequency standards, 
optical magnetometry and 
few-photon nonlinear optics.

Consider the propagation of four nearly resonant plane 
waves, parallel or anti-parallel to the $z$ axis, in a 
medium consisting of double-$\Lambda$ atoms (see Fig.~1).
These include two counter-propagating 
driving fields with equal frequencies 
$\nu_{d}$ and (complex) 
Rabi-frequencies $E_f$ and $E_b$, 
and two generated fields (anti-Stokes and Stokes)  with carrier frequencies 
$\nu_1$ and $\nu_2$ obeying $\nu_1+\nu_2=2\nu_d$.
The fields interact via the long-living coherence (decay rate 
$\gamma_0$) on the
transition between the  ground state sub-levels 
$b_1$ and $b_2$ with frequency splitting  $\omega_0=\omega_{b2}-
\omega_{b1}$.

Due to resonantly enhanced four-wave mixing 
the coherent pump fields $E_f$ and $E_b$
generate counter-propagating anti-Stokes and Stokes fields 
(here described by the
complex Rabi-frequencies $E_1$ and $E_2$). For a sufficiently large
density-length product of the medium and for a certain 
pump field intensity, the system shows a phase-transition
to self-oscillations \cite{zibrov98}. 
The feedback mechanism required for oscillations
is provided here by the 
gain medium itself: A spontaneously generated Stokes 
photon stimulates ``downstream''  
a Raman process. As 
a result an anti-Stokes photon is generated with a fixed relative phase. 
This photon propagates in the opposite direction and
stimulates another scattering event ``upstream''. 
If the phase matching condition
is fulfilled, this causes a second Stokes emission 
in phase with the first one closing the feedback cycle. 
A crucial condition for the coherence of this feedback mechanism is 
a sufficiently long-lived Raman coherence.

We now discuss the transition to self-oscillation and the
classical and quantum dynamics of the oscillator above threshold in detail.
To simplify the analysis we ignore inhomogeneous broadening and
assume equal coupling strength of all four fields as well as equal 
radiative decays out of the upper levels. Furthermore we assume
that the forward driving field $E_f$ is in resonance with the 
$b_2\to a_1$ transition, whereas the backward driving field
$E_b$ has a detuning $\Delta\gg |E_b|$ from the 
$b_1\to a_2$ transition. In this case linear losses of the fields 
due to single-photon absorption processes are minimal.

In order to calculate the medium response
we solve the single-atom density matrix equations
in third order of the Stokes and anti-Stokes fields  and 
assume $|\Delta| \gg \gamma,|E_{f,b}| \gg \gamma_0,\delta$,
where $\delta=\nu_d+\omega_0-\nu_1$ is the two-photon detuning.
In a frame rotating with the carrier frequencies,
the propagation of the classical 
fields can then be described by the following
equations for the slowly varying complex Rabi-frequencies
\begin{eqnarray}
\frac{\rm d}{{\rm d}z} {\widetilde E}_1 &=& 
 i\kappa \frac{{\widetilde E}_1^2 E_2 E_f^*{\widetilde E}_b^*+
E_f{\widetilde E}_b E_2^*(2|{\widetilde E}_1|^2 -|E_f|^2)}{\Delta
|E_f|^4}\nonumber\\
&&+i\left[\kappa
\frac{i\gamma_0-\delta}
{|E_f|^2} -\kappa \frac{|{\widetilde E}_b|^2-|E_f|^2}{\Delta|E_f|^2}-\Delta k\right] 
{\widetilde E}_1, \label{E_1}\\
\frac{\rm d}{{\rm d}z} E_2^* &=& i\kappa
\frac{(|{\widetilde E}_1|^2 -|E_f|^2)E_f^*{\widetilde E}_b^* 
{\widetilde E}_1}{\Delta|E_f|^4},\label{E_2}\\
\frac{\rm d}{{\rm d}z} E_f &=& i\kappa \frac{{\widetilde E}_1^* E_2^* 
{\widetilde E}_b E_f^2}{\Delta |E_f|^4},\\
\frac{\rm d}{{\rm d}z} {\widetilde E}_b^* &=& 
-i\kappa \frac{{\widetilde E}_1^* E_2^*E_f|E_f|^2}
{\Delta|E_f|^4}.\label{O_2}
\end{eqnarray}
In these equations we have kept ac-Stark induced 
phase terms only in lowest order
of the generated fields, since we are interested in  
in the case $|E_{f,b}|  \gg |E_{1,2}|$.
$\kappa = (3/8\pi) N\lambda^2 \gamma_a$ is the equal coupling constant
of all fields with $N$ being the atom density, $\lambda$ the average
wavelength of the fields, and $\gamma_a$ the common population decay rate
out of the excited states.
$E_1={\widetilde E}_1 {\rm e}^{i(\Delta k -\kappa/\Delta)z}$,
$E_b={\widetilde E}_b {\rm e}^{i\kappa z/\Delta}$, with 
$\Delta k=
k_2-k_1$ 
being the phase mismatch. 

We note an important feature of
the nonlinear coupling in Eqs. (\ref{E_1}) and (\ref{E_2}): In contrast to
usual $\chi^{(3)}$-media, the lowest-order cross-coupling terms 
are proportional to the ratio
of the pump fields 
rather than the product. 
\begin{equation}
\frac{\rm d}{{\rm d}z} E_1\sim - i\chi^{(3)}\, E_fE_b\,  E_2^*
\enspace \longrightarrow\enspace
 - i\frac{\kappa}{\Delta}
\, \frac{E_b}{E_f^*}\, E_2^*.
\end{equation}
Thus for $|E_f|=|E_b|$
these terms are independent of the pump-field amplitudes. 
We will
see later on that this leads to a rather unusual threshold behavior.

In the present system a transition to spontaneous oscillations
is possible \cite{zibrov98}, if the phase matching condition
\begin{equation}
\kappa
\frac{\delta}
{|E_f|^2} +\kappa \frac{|E_b|^2-|E_f|^2}
{\Delta|E_f|^2}+\Delta k
=0
\label{match}
\end{equation}
is fulfilled.
For large values of $\kappa$, Eq.(\ref{match}) 
describes a  pulling of the frequency differences between 
generated fields and driving fields 
towards the ac-Stark shifted frequency of the Raman transition
\begin{eqnarray}
\nu_1-\nu_{d1}=\nu_{d2}-\nu_2 &=& \frac{\eta\,    \Bigl[\omega_0 + 
(|E_b|^2-|E_f|^2)/\Delta\Bigr]}{
1+\eta}
.\label{locking}
\end{eqnarray}
This equation shows a close analogy with intracavity EIT. 
$\eta=c\kappa/2|E_f|^2$ 
is a frequency stabilisation factor
\cite{lukin98ol}. This factor also governs the 
group velocity of the eigenmodes of the system $v_{gr}=c/(1+\eta)$
and can be rather large. 
 For conditions close to the experiments of 
Ref.\cite{zibrov98}, a reduction factor of $\eta\sim 5\times 10^6$   
was  measured \cite{group_delay_preprint}. 
In the limit of large $\eta$ the beat-notes between generated and pump fields 
locks tightly to the Raman-transition frequency of the medium.

We next consider the classical steady state solution of 
the  propagation problem. In the ideal limit  ($\gamma_0=0$)
Eqs.(\ref{E_1}-\ref{O_2}) have 
four constants of motion:  the
total intensity of the generated and  pump fields $|E_1|^2+|E_2|^2$
and $|E_f|^2+|E_b|^2$, as well as ${\rm Re}\, [E_f^*E_b^*
E_1 E_2]$ which has a similar 
structure to the cubic expression conserved in 3-wave mixing
\cite{Armstrong62}.
There is also the somewhat unusual 
constant of motion, $|E_f|^2\exp\bigl(|E_1|^2/|E_f|^2
\bigr)$ \cite{referee}. If we take into account, however, that 
Eqs.~(\ref{E_1}-\ref{O_2}) only hold to third order in the generated fields,
this constant is equivalent to $|E_f|^2+|E_1|^2$.
With this Eqs.(\ref{E_1}-\ref{O_2})
can be solved analytically, if the  phase matching condition is 
approximately fulfilled. 
Assuming equal input intensities of the driving fields 
$|E_f(0)|=|E_b(L)|$
at $z=0$ and 
$z=L$ respectively ($L$ being the cell length), 
and disregarding linear losses 
due to the finite lifetime of the ground-state 
coherence, one finds in second order of the generated fields
\begin{eqnarray}
|E_1(z)| &=& E\, \sin\vartheta(z),\qquad |E_2(z)| = E\, \cos\vartheta(z), \\
|E_f(z)| &=& \Bigl[|E_f(0)|^2 -E^2\, \sin^2\vartheta(z)\Bigr]^{1/2},\\
|E_b(z)| &=& \Bigl[|E_f(0)|^2 -E^2\, \cos^2\vartheta(z)\Bigr]^{1/2},
\end{eqnarray}
where $\vartheta(z) =  \kappa z/\Delta\left(1-
E^2/2|E_f(0)|^2\right)$. For $\kappa L/\Delta <\pi/2$, $E\equiv 0$.
For values of 
$\kappa L/\Delta$ larger than the critical value $\pi/2$ there
are two solutions
\begin{eqnarray}
E=\sqrt{2}\, |E_f(0)| \, 
\cases{ 0 \cr
 \sqrt{1-\frac{\pi}{2}\frac{\Delta}{\kappa L}} 
}
\qquad{\rm for}\qquad \frac{\kappa L}{\Delta} \ge\frac{\pi}{2}
\end{eqnarray}
with $E= 0$ being unstable. 
It should be noted that in contrast to degenerate 4-wave mixing 
in usual $\chi^{(3)}$-media \cite{Yuen79}, the threshold condition does not
contain the amplitude of the pump fields. Thus under the ideal conditions 
assumed here, i.e. for an infinitely long lived dark state, an arbitrarily
small stationary pump intensity is sufficient 
to reach the oscillation threshold. 
Fig.~2 shows the above-threshold
behavior of $E$ as a function of $\kappa L/\Delta $ and the  
field amplitudes normalized to $|E_f(0)|$ inside the cell.
If the system oscillates not too far above threshold, the depletion 
of the  pump fields is small and we may assume in the
following constant driving field 
amplitudes, $|E_f(z)|=|E_b(z)|
=E_d$.

To calculate the linewidth  of  Stokes and 
anti-Stokes fields relative to the drive field above threshold,
we assume that the generated fields  can be represented as a 
sum of the classical stationary solutions  and a time-dependent
fluctuation  ($\hat E_{1,2}(z,t)= E_{1,2}(z)  + 
\delta E_{1,2}(z,t) $).
We utilize 
a standard linearized c-number Langevin approach
in which collective atomic variables and fields
are described by  
time- and position-
dependent stochastic differential equations
with $\delta-$correlated 
Langevin forces\cite{fleischhauer95pra}. 
 The 
diffusion coefficients 
or noise correlations are derived using the 
fluctuation-dissipation theorem and
generalized Einstein relations.
 We obtain for the 
 Fourier-components of the Stokes and anti-Stokes
fluctuations $\delta E_{1,2}(\omega)\equiv 1/\sqrt{2 \pi}\int{\rm d}t
\, \delta E_{1,2}(t)\, {\rm e}^{-i\omega t}$
\begin{equation} \label{max} 
\frac{d}{d z}\left[\matrix{\delta E_1^* \cr \delta E_2}\right]
= i\left[\matrix{
-{\kappa\omega/E_d^2}-{\omega/c} & 
{\kappa/\Delta}\cr 
{\kappa/\Delta} & 
{\omega/c}}\right]\left[\matrix
{\delta E_1^*\cr \delta E_2}\right]
+\left[\matrix{f_1^*\cr f_2}\right],
\end{equation}
where small fluctuation frequencies ($\omega\ll E_d $) 
and a constant phase of the pump field have been assumed.
The  term $-\kappa\omega/E_d ^2$ is the contribution from the
medium dispersion. Following the procedure of 
Ref.\cite{fleischhauer95pra},
we find for the dominant noise correlation
\begin{eqnarray} 
\langle f_1(z,\omega) f_2(z',\omega') \rangle &\simeq&   
{\kappa^2 L \over {\cal N}} { i \over \Delta} 
\delta(z-z') \delta(\omega+\omega'),
\label{noise}
\end{eqnarray}
where ${\cal N}$ is the number of atoms in the cell, and
we have identified the quantization length defined in 
\cite{fleischhauer95pra}
with the length of the cell $L$. 

Solving the inhomogeneous boundary problem for the
Fourier-components of the field
fluctuations with $\delta E_1^*(0,\omega)=0$ and $\delta E_2(L,\omega)=0$,
we eventually find for the phase 
of e.g. $\delta E_1$
\begin{eqnarray}
\label{sol1}
&&\delta\phi_1(L,\omega) = -\frac{i\kappa c}{\Delta |E_1(L)|}\nonumber\\
&&\times\int_0^L\!\!{\rm d} z
\frac{{\rm Im}[f_1(z)]\cos\vartheta(z)
+{\rm Re}[f_2(z)] \sin\vartheta(z)}
{\omega \left(1+ \eta\right)}.
\end{eqnarray}

In phase-diffusion approximation, the linewidth $\Delta\nu_1$ of  $E_1$
relative to the pump field is given by
\begin{equation} 
\label{phdif2}
\langle\delta\phi_1(L,\omega) \delta\phi_1(L,\omega ') \rangle 
=\Delta\nu_1\ \frac{\delta(\omega+\omega^\prime)}{\omega^2}.
\end{equation}
Using Eqs.(\ref{noise}) and (\ref{sol1}) we  arrive at 
\begin{eqnarray} 
 \Delta \nu_1 =
  \frac{2 E_d ^4}{\Delta^2} 
\frac{\hbar \nu }{P_{out}},   
\label{linew1}
\end{eqnarray}    
where $P_{out}$ is the output power of the  mode. 

Eq.(\ref{linew1}) can be represented in a very instructive form, if 
the group-delay time $\tau_{gr}= L/c (1+\eta)$ in the resonant medium 
is introduced. In the appropriate 
limit, $\eta\gg 1$, and near threshold
such that $\kappa L/\Delta\approx \pi/2$
the linewidth can be written as 
\begin{equation}
\Delta\nu_1 = \frac{\pi^2}{8} \, \tau_{gr}^{-2}\, 
\frac{\hbar \nu }{P_{out}}\approx  \tau_{gr}^{-2}\, 
\frac{\hbar \nu }{P_{out}}.
\label{dbu}
\end{equation}
Eq.(\ref{dbu}) is formally
identical to that of an ideal laser with the cavity decay time
replaced by the group delay time. In usual 4-wave mixing,
 based on non-resonant
Kerr-nonlinearities \cite{old1,old2,old3,old4}, 
the group velocity is essentially equal
to the vacuum speed of light. 
In the present scheme, however, it can be substantially reduced
due to EIT.

It is important to emphasize that the photon pairs generated by the 
oscillation process near threshold  
are in quantum-mechanically correlated states. This results 
in a dramatic suppression of intrinsic quantum fluctuations in a 
quadrature of the combined mode \cite{lukin99}.

In the discussion above we have neglected the relaxation
rate of the ground state coherence $\gamma_0$.
If this decay is taken into account, one finds the modified 
threshold condition:
$ {\rm cos}(\xi L) + (\gamma_{0} \Delta)/(2 E_d ^2) 
{\rm sin}(\xi L) =0,
$
with $\xi = \kappa \sqrt{1/\Delta^2- 
\gamma_0^2/4 E_d ^4}$. In particular, oscillation can be achieved only
if $E_d ^2\ge \gamma_0 |\Delta|/2$. This can be translated
into a condition for the photon flux $\Phi$, i.e. 
the number of pump-photons traversing the cell per unit time.
One finds that the threshold photon flux in each pump beam is equal
to the  number of atoms in the ensemble 
decaying out of the dark
state per unit time:
\begin{equation}
\Phi_{\rm th}=f\, {\cal N} \, \gamma_0,\label{n_thresh}
\end{equation}
where $f$ is a numerical prefactor of order unity.
Since by using buffer gases or  coated cells very small values
of $\gamma_0$ can be achieved, a threshold flux corresponding to only 
few photons in the cell is feasible, leading to an interesting new regime of 
nonlinear optics.

Furthermore, the non-vanishing linear
losses resulting from the decay of the ground-state coherence lead to an
additional noise contribution to the linewidth
\begin{eqnarray} 
\Delta \nu_1 =
\frac{\pi^2}{8} \, \tau_{gr}^{-1}
\left(\tau_{gr}^{-1}+ 2\gamma_0\right)
\, 
\frac{\hbar \nu }{P_{out}}.
\label{linew2}
\end{eqnarray} 
This result can easily be interpreted. $P_{out} \tau_{gr} /\hbar\nu$
is equal to twice the  number of  Stokes or anti-Stokes photons
in the cell. (Note that the total number of Stokes plus
anti-Stokes photons is spatially homogeneous.) Like in a usual
laser, photon correlations are maintained over a time equal to the
number of photons multiplied by the time a single photon
stays in the system \cite{Kuppens94}. 
The latter time is here given by the group delay time
(if $\gamma_0$ is sufficiently small). If the lifetime
of the dark state becomes shorter than the group delay, the phase
information carried by a photon is lost faster and $\tau_{gr}^{-1}$
is dominated by $2\gamma_0$.
Thus the  minimum
linewidth is ultimately determined by the lower-level coherence decay.

Similar to  the case discussed in Ref. \cite{lukin98ol} for the 
intracavity system, 
the present result for the frequency locking, Eq.(\ref{locking}), and the 
linewidth, Eqs.(\ref{linew1},\ref{linew2}) are a consequence of the large 
atomic dispersion associated with two-photon resonances in
phase coherent media. In the limit
of long-lived ground-state coherences, the beat-note linewidth can 
be extremely narrow. At the same time the resonantly enhanced nonlinearity
makes it possible to achieve oscillation with very low pump powers.

In order to see, whether the small intrinsic linewidth can indeed
be exploited, we now estimate the influence of systematic effects 
on the beat-note
frequency. 
The most serious limitations  arise from the ac-Stark shifts as indicated by 
Eq.(\ref{locking}).  At large values of pump intensities these shifts 
are large and hence fluctuations in pump powers and frequencies 
will result in associated broadening of the oscillator linewidth. 
However, the resonantly enhanced nonlinearity makes oscillation 
possible already when $E_d ^2\ge \gamma_0\Delta$, i.e. when the
near-resonant ac-Stark shift  $E_d ^2/\Delta$ exceeds the
ground-state coherence decay $\gamma_0$. Thus with
stabilized pump frequencies and intensities, technical fluctuations of 
the beat-frequency due to ac-Stark shifts 
could be several orders of magnitude smaller than $\gamma_0$. In the 
experiment of Ref.\cite{zibrov98}, for instance, short-term linewidth values  
below $100$ Hz have been measured even though the transient time broadening
of a Raman transition was about $50$ kHz. It is clear that 
observation of quantum-limited linewidth of the oscillator is most 
likely in the regime of ultra-low pump intensities.  It is however this 
regime which is of main interest for quantum control and manipulation of 
quantum properties of few photon fields \cite{lukin99}.

In conclusion, we have demonstrated that resonant nonlinear
interactions involving atomic coherence can be used for efficient 
generation of non-classical photon fields 
with a stable and narrow beat-note linewidth
and small pump requirements.
We expect these features to be of interest 
in many areas of quantum and nonlinear optics.

The authors gratefully acknowledge useful discussions with 
P.~Hemmer, V.~Sautenkov,  and 
A.~Zibrov and the support from the National Science Foundation, 
the Office of Naval Research and the Welch Foundation. 
M. F. would like to thank the Institute for Theoretical Atomic and
Molecular Physics at the Harvard-Smithsonian Institution 
for the hospitality during his stay.



\begin{figure}

\

\epsfxsize=4.5cm
\centerline{\epsffile{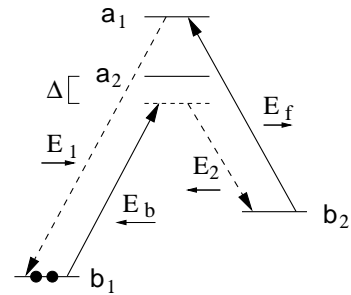}}

\

\caption{Atoms in double $\Lambda$ configuration interacting 
with two classical driving fields in forward ($E_f$) and
backward direction
($E_{b}$) and two quantum fields ($E_{1,2}$). All optical transitions
are assumed to be radiatively broadened.}

 \end{figure}
\begin{figure}

\

\epsfxsize=8.5 cm
\centerline{\epsffile{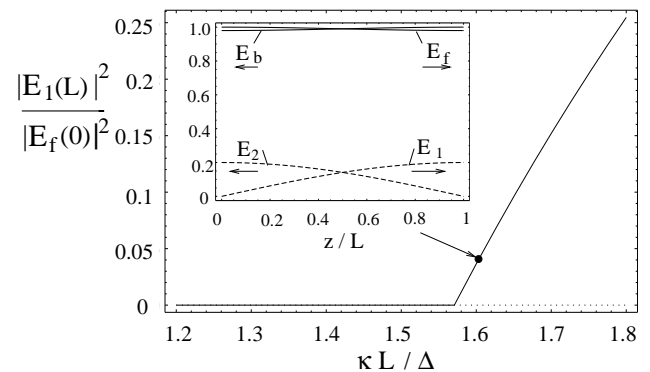}}

\

\caption{Phase transition to mirrorless parametric oscillations.
Analytic solution for amplitude of 
generated field for $\gamma_0=0$.
Insert: Normalized field amplitudes inside medium for $E/|E_f(0)|=0.2$
}
 \end{figure}

\end{document}